# Theoretical study of polarization tracking in satellite quantum key distribution

Jing Ma [1], Guangyu Zhang [1, 2], Yiwen Rong [1], Liying Tan [1]

1. National Key Laboratory of Tunable Laser Technology, Harbin Institute of Technology, Harbin 150001, China
2. College of Applied Sciences, Harbin University of Science and Technology, Harbin 150080, China

**Abstract:** The BB84 and B92 protocols based on polarization coding are usually used in free-space quantum key distribution. Polarization tracking technique is one of the key techniques in the satellite quantum key distribution. Because the photon polarization state will be changed as a result of the satellite movement, both the transmitter and receiver need to have the ability to track the polarization orientation variation to accomplish the quantum cryptography protocols. In this paper, the polarization tracking principles are analyzed based on Faraday effect and the half-wave plate. The transforms of six photon polarization states in three conjugative bases are given and the quantum key coding principles based on the polarization tracking are analyzed for the BB84 and B92 protocols.

## 1 Introduction

Quantum cryptography, or more accurately, quantum key distribution (QKD), uses the single-photon optical communication technique to distribute the secret keys required for confidential communications, and its security is guaranteed by the Heisenberg's uncertainty principle and the quantum no-cloning principle.

Free-space QKD has achieved remarkable progresses since Bennett et al demonstrated the first QKD experiment [1]. Current advances are mainly focusing on demonstrations of QKD over the point-to-point atmospheric optical paths [2~6] and researches of feasibility on the ground-to-satellite, satellite-to-ground and satellite-to-satellite QKD [7,8]. In the point-to-point experiments that have already been carried out, the BB84 and B92 protocols based on polarization coding were usually used [9,10]. In order to realize QKD between any two arbitrary locations on the globe, a satellite is needed to be used as a secure relay station. Polarization tracking technique is one of the key techniques in the satellite QKD. Because the photon polarization state will be changed as a result of the satellite movement, both the transmitter and receiver need to have the ability to track the polarization orientation variation to accomplish the quantum cryptography protocols. Nordholt et al have already presented to use the half-wave plate for the polarization tracking [7], but the relevant theoretical analysis was not given.

In this paper, the polarization tracking principles based on Faraday effect and the half-wave plate are first analyzed, then the transforms of six photon polarization states in three conjugative bases are given, and at last the quantum key coding principles based on the polarization tracking are analyzed for the BB84 and B92 protocols.

## 2 Polarization tracking principle

In the satellite QKD, the change of the photon polarization orientation can be perceived as doing an operation on the photon polarization state and it can be expressed by the operators. Generally the polarization tracking can be expressed as

$$F_1 F_2 |\psi\rangle = |\psi\rangle \tag{1}$$

where $|\psi\rangle$ is the photon polarization state, and $F_1$ and $F_2$ are two operators. $F_2$ represents the change of the photon polarization state, and $F_1$ represents the compensation for the polarization orientation variation.

### 2.1 Based on Faraday effect

The polarization tracking can be realized based on Faraday effect in the satellite QKD. Faraday effect is the effect that an optical activity happens under strong magnetic field to some mediums. Under the action of magnetic field, some mediums gain the optical activity and rotate the polarization plane of the incident light. The angle $\beta$ that the incident light is rotated is proportional to the length of the medium $l$ and the magnetic induction intensity $B$, that is

$$\beta = VBl \tag{2}$$

where V is the Verdet constant which depends on the characters of the medium. According to Eq. (2), there is

$$B = \frac{\beta}{Vl} \tag{3}$$

During the process of QKD, both the transmitter and receiver have to determine the polarization base, that is, a polarization "zero" direction that can be defined by a beacon from either the transmitter or the receiver. In the polarization tracking, the rotation angle of the incident light can be controlled by changing the magnetic induction intensity, and the polarization "zero" direction can be compensated.

Assuming that the "zero" direction polarization state is noted as $|0\rangle$, it changes the angle $\theta$, represented by the operator $D$, and there is

$$D|0\rangle = |\theta\rangle \tag{4}$$

Faraday effect is applied for the polarization tracking and its effect can be noted by the operator $F$. There is

$$FD|0\rangle = F|\theta\rangle = |0\rangle \tag{5}$$

From above, it is seen that the polarization "zero" direction has been compensated.

**2.2 Based on the half-wave plate**

For the satellite QKD, the polarization tracking can be also realized by using the half-wave plate. Here, the negative uniaxial crystal is considered for example. Assuming that the oscillating plane of the incident light and the axis form the angle $\alpha$, the oscillating plane rotates the angle $2\alpha$ after the light transmits through the crystal. If the incident light is the right-circular polarization, it will turn to the left-circular polarization and vice versa.

In the polarization tracking, the polarization "zero" direction can be corrected by rotation of the half-wave plate. The effect of the half-wave plate can be noted as the operator $H_{\lambda/2}$. Assuming that the axis of the half-wave plate and the polarization "zero" direction form the angle $\theta/2$, and there is

$$H_{\lambda/2}D|0\rangle = H_{\lambda/2}|\theta\rangle = |0\rangle \tag{6}$$

From above, the polarization "zero" direction has been compensated.

## 3 The transformation of photon polarization state

For the BB84 and B92 protocols based on polarization coding, six photon polarization states are involved: horizontal polarization, vertical polarization, $45°$ polarization, $135°$ polarization, left-circular polarization and right-circular polarization. They can be noted as $|H\rangle$, $|V\rangle$, $|\pi/4\rangle$, $|3\pi/4\rangle$, $|L\rangle$ and $|R\rangle$. They belong to three conjugative bases: $\{|H\rangle, |V\rangle\}$, $\{|\pi/4\rangle, |3\pi/4\rangle\}$ and $\{|L\rangle, |R\rangle\}$.

Assuming that the horizontal polarization state is set as the polarization "zero" direction, that is

$$|H\rangle = |0\rangle \tag{7}$$

If Faraday effect is applied for the polarization tracking, for an arbitrary linear

polarization state $|\varphi\rangle$, there is

$$D|\varphi\rangle = |\varphi + \theta\rangle \tag{8}$$

$$FD|\varphi\rangle = F|\varphi + \theta\rangle = |\varphi\rangle \tag{9}$$

For the circular polarization state, there is

$$D|L\rangle = |L\rangle \tag{10}$$

$$D|R\rangle = |R\rangle \tag{11}$$

$$FD|L\rangle = |L\rangle \tag{12}$$

$$FD|R\rangle = |R\rangle \tag{13}$$

If the half-wave plate is used for the polarization tracking, for the linear polarization state, the effect of the half-wave plate can be regard as the symmetric operation, transforming the original direction to the mirror image, regarding the axis as the symmetry axis. Assuming that the axis lies in the angle $\theta/2$, and there is

$$H_{\lambda/2}D|\varphi\rangle = H_{\lambda/2}|\varphi + \theta\rangle = |-\varphi\rangle \tag{14}$$

According to Eq. (14), there is

$$H_{\lambda/2}D|H\rangle = H_{\lambda/2}D|0\rangle = H_{\lambda/2}|\theta\rangle = |0\rangle = |H\rangle \tag{15}$$

$$H_{\lambda/2}D|V\rangle = H_{\lambda/2}D\left|\frac{\pi}{2}\right\rangle = H_{\lambda/2}\left|\frac{\pi}{2} + \theta\right\rangle = \left|-\frac{\pi}{2}\right\rangle = \left|\frac{\pi}{2}\right\rangle = |V\rangle \tag{16}$$

$$H_{\lambda/2}D\left|\frac{\pi}{4}\right\rangle = H_{\lambda/2}\left|\frac{\pi}{4} + \theta\right\rangle = \left|-\frac{\pi}{4}\right\rangle = \left|\frac{3}{4}\pi\right\rangle \tag{17}$$

$$H_{\lambda/2}D\left|\frac{3}{4}\pi\right\rangle = H_{\lambda/2}\left|\frac{3}{4}\pi + \theta\right\rangle = \left|-\frac{3}{4}\pi\right\rangle = \left|\frac{\pi}{4}\right\rangle \tag{18}$$

For the circular polarization state, there is

$$H_{\lambda/2}D|L\rangle = H_{\lambda/2}|L\rangle = |R\rangle \tag{19}$$

$$H_{\lambda/2}D|R\rangle = H_{\lambda/2}|R\rangle = |L\rangle \tag{20}$$

It can be seen from above that the operator $D$ does not affect $|L\rangle$ and $|R\rangle$. When Faraday effect is applied, the linear and circular polarization states have returned to the original states after being acted by the operator $D$ and $F$. When the half-wave plate is used, $|H\rangle$ and $|V\rangle$ have returned to the original states after being acted by the operator $D$ and $H_{\lambda/2}$, but $|\pi/4\rangle$, $|3\pi/4\rangle$, $|L\rangle$ and $|R\rangle$ have changed.

## 4 Quantum key coding principle based on polarization tracking

In the satellite QKD, the BB84 and B92 protocols based on the polarization coding are used. The BB84 protocol is based on four nonorthogonal states belonging to two conjugative bases and two states in each basis are orthogonal. Two conjugative bases can be chosen from $\{|H\rangle,|V\rangle\}$, $\{|\pi/4\rangle,|3\pi/4\rangle\}$ and $\{|L\rangle,|R\rangle\}$. The B92 protocol is based on any two nonorthogonal states that can be chosen from $|H\rangle$, $|V\rangle$, $|\pi/4\rangle$, $|3\pi/4\rangle$, $|L\rangle$ and $|R\rangle$. In the situation based on Faraday effect, the quantum key coding principle is same as the BB84 and B92 protocols because all photon polarization states can be compensated to the original states. The following part mainly focuses on the analysis based on the half-wave plate.

For the BB84 protocol, if the transmitter uses $|H\rangle$ and $|V\rangle$ for coding, $|H\rangle$ represents the binary number "1" and $|V\rangle$ represents "0", that is

$$|H\rangle \leftrightarrow 1 \tag{21}$$

$$|V\rangle \leftrightarrow 0 \tag{22}$$

After being operated by $D$ and $H_{\lambda/2}$, they are projected on $\{|H\rangle,|V\rangle\}$, and there is

$$P_H H_{\lambda/2} D|H\rangle = P_H|H\rangle = |H\rangle\langle H|H\rangle = |H\rangle \tag{23}$$

$$P_V H_{\lambda/2} D|V\rangle = P_V|V\rangle = |V\rangle\langle V|V\rangle = |V\rangle \tag{24}$$

where $P_H$ and $P_V$ are the projection operators. So the receiver's coding is

$$|H\rangle \leftrightarrow 1 \tag{25}$$

$$|V\rangle \leftrightarrow 0 \tag{26}$$

If the transmitter uses $|\pi/4\rangle$ and $|3\pi/4\rangle$ for coding, that is

$$\left|\frac{\pi}{4}\right\rangle \leftrightarrow 1 \tag{27}$$

$$\left|\frac{3}{4}\pi\right\rangle \leftrightarrow 0 \tag{28}$$

After being operated by $D$ and $H_{\lambda/2}$, they are projected on $\{|\pi/4\rangle,|3\pi/4\rangle\}$, and there is

$$P_{3\pi/4}H_{\lambda/2}D\left|\frac{\pi}{4}\right\rangle = P_{3\pi/4}\left|\frac{3}{4}\pi\right\rangle = \left|\frac{3}{4}\pi\right\rangle\left\langle\frac{3}{4}\pi\bigg|\frac{3}{4}\pi\right\rangle = \left|\frac{3}{4}\pi\right\rangle \tag{29}$$

$$P_{\pi/4}H_{\lambda/2}D\left|\frac{3}{4}\pi\right\rangle = P_{\pi/4}\left|\frac{\pi}{4}\right\rangle = \left|\frac{\pi}{4}\right\rangle\left\langle\frac{\pi}{4}\bigg|\frac{\pi}{4}\right\rangle = \left|\frac{\pi}{4}\right\rangle \tag{30}$$

where $P_{3\pi/4}$ and $P_{\pi/4}$ are the projection operators    So the receiver's coding is

$$\left|\frac{3}{4}\pi\right\rangle \leftrightarrow 1 \tag{31}$$

$$\left|\frac{\pi}{4}\right\rangle \leftrightarrow 0 \tag{32}$$

If the transmitter uses $|L\rangle$ and $|R\rangle$ for coding, that is

$$|L\rangle \leftrightarrow 1 \tag{33}$$

$$|R\rangle \leftrightarrow 0 \tag{34}$$

After being operated by $D$ and $H_{\lambda/2}$, they are projected on $\{|L\rangle, |R\rangle\}$, and there is

$$P_R H_{\lambda/2} D|L\rangle = P_R|R\rangle = |R\rangle\langle R|R\rangle = |R\rangle \tag{35}$$

$$P_L H_{\lambda/2} D|R\rangle = P_L|L\rangle = |L\rangle\langle L|L\rangle = |L\rangle \tag{36}$$

where $P_R$ and $P_L$ are the projection operators. So the receiver's coding is

$$|R\rangle \leftrightarrow 1 \tag{37}$$

$$|L\rangle \leftrightarrow 0 \tag{38}$$

For the B92 protocol, if the transmitter uses $|H\rangle$ and $|\pi/4\rangle$ for coding, that is

$$|H\rangle \leftrightarrow 1 \tag{39}$$

$$\left|\frac{\pi}{4}\right\rangle \leftrightarrow 0 \tag{40}$$

After being operated by $D$ and $H_{\lambda/2}$, they are projected on $|V\rangle$ and $|\pi/4\rangle$, and there is

$$P_{\pi/4}H_{\lambda/2}D|H\rangle = P_{\pi/4}|H\rangle = \left|\frac{\pi}{4}\right\rangle\left\langle\frac{\pi}{4}\bigg|H\right\rangle = \left|\frac{\pi}{4}\right\rangle/\sqrt{2} \tag{41}$$

$$P_V H_{\lambda/2} D \left|\frac{\pi}{4}\right\rangle = P_V \left|\frac{3}{4}\pi\right\rangle = |V\rangle\langle V|\frac{3}{4}\pi\rangle = |V\rangle/\sqrt{2} \tag{42}$$

So the receiver's coding is

$$\left|\frac{\pi}{4}\right\rangle \leftrightarrow 1 \tag{43}$$

$$|V\rangle \leftrightarrow 0 \tag{44}$$

If the transmitter uses $|H\rangle$ and $|L\rangle$ for coding, that is

$$|H\rangle \leftrightarrow 1 \tag{45}$$

$$|L\rangle \leftrightarrow 0 \tag{46}$$

After being operated by $D$ and $H_{\lambda/2}$, they are projected on $|V\rangle$ and $|L\rangle$, and there is

$$P_L H_{\lambda/2} D |H\rangle = P_L |H\rangle = |L\rangle\langle L|H\rangle = |L\rangle/\sqrt{2} \tag{47}$$

$$P_V H_{\lambda/2} D |L\rangle = P_V |R\rangle = |V\rangle\langle V|R\rangle = |V\rangle/\sqrt{2} \tag{48}$$

So the receiver's coding is

$$|L\rangle \leftrightarrow 1 \tag{49}$$

$$|V\rangle \leftrightarrow 0 \tag{50}$$

If the transmitter uses $|\pi/4\rangle$ and $|L\rangle$ for coding, that is

$$\left|\frac{\pi}{4}\right\rangle \leftrightarrow 1 \tag{51}$$

$$|L\rangle \leftrightarrow 0 \tag{52}$$

After being operated by $D$ and $H_{\lambda/2}$, they are projected on $|\pi/4\rangle$ and $|L\rangle$, and there is

$$P_L H_{\lambda/2} D \left|\frac{\pi}{4}\right\rangle = P_L \left|\frac{3}{4}\pi\right\rangle = |L\rangle\langle L|\frac{3}{4}\pi\rangle = |L\rangle/\sqrt{2} \tag{53}$$

$$P_{\pi/4} H_{\lambda/2} D |L\rangle = P_{\pi/4} |R\rangle = \left|\frac{\pi}{4}\right\rangle\langle\frac{\pi}{4}|R\rangle = \left|\frac{\pi}{4}\right\rangle/\sqrt{2} \tag{54}$$

So the receiver's coding is

$$|L\rangle \leftrightarrow 1 \tag{55}$$

$$\left|\frac{\pi}{4}\right\rangle \leftrightarrow 0 \qquad (56)$$

From the analysis above, it can be seen that the quantum key coding principle is the same as the BB84 and B92 protocols for $\{|H\rangle,|V\rangle\}$ when the half-wave plate is used for the polarization tracking. But because $|\pi/4\rangle$, $|3\pi/4\rangle$, $|L\rangle$ and $|R\rangle$ have changed after being operated by $D$ and $H_{\lambda/2}$, the receiver has to adjust the measuring bases to accomplish the BB84 and B92 protocols. So the quantum key coding principle is different from the BB84 and B92 protocols for $\{|\pi/4\rangle,|3\pi/4\rangle\}$ and $\{|L\rangle,|R\rangle\}$.

## 5 Conclusion

In this paper, the polarization tracking are presented based on Faraday effect. The polarization tracking principles are analyzed based on Faraday effect and the half-wave plate for the satellite QKD. The polarization "zero" direction can be corrected by adjusting the magnetic induction intensity and by the rotation of the half-wave plate. By analyzing the transformation relations of the six photon polarization states in three conjugative bases, the quantum key coding principles based on the polarization tracking are analyzed and the BB84 and B92 protocols can finally be accomplished.